\documentclass[10pt, reqno, letterpaper, twoside]{amsart}
\usepackage[margin=1in]{geometry}

\usepackage{amssymb, bm, mathtools}
\usepackage[usenames,dvipsnames,svgnames,table]{xcolor}
\usepackage[pdftex, xetex]{graphicx}
\usepackage{enumerate, setspace}
\usepackage{csquotes}

\usepackage{float, colortbl, tabularx, longtable, multirow, subcaption, environ, wrapfig, textcomp, booktabs}
\usepackage{pgf, tikz, framed, url, hyperref}
\usepackage[normalem]{ulem}

\usepackage[backend=biber, style=numeric, defernumbers=true, sorting=none]{biblatex}
\usetikzlibrary{arrows,positioning,automata,shadows,fit,shapes}
\usepackage{microtype}
\microtypecontext{spacing=nonfrench}

\usepackage[english]{babel}

\usepackage{times}
\title{GANet-Seg: Adversarial Learning for Brain Tumor Segmentation with Hybrid Generative Models}
\author{
Qifei Cui* [qifei@seas],
Xinyu Lu [xinyulu@seas],
}

\begin{document}
\begin{refsection}
\maketitle

\begin{abstract}
This work introduces a novel framework for brain tumor segmentation leveraging pre-trained GANs and Unet architectures. By combining a global anomaly detection module with a refined mask generation network, the proposed model accurately identifies tumor-sensitive regions and iteratively enhances segmentation precision using adversarial loss constraints. Multi-modal MRI data and synthetic image augmentation are employed to improve robustness and address the challenge of limited annotated datasets. Experimental results on the BraTS dataset demonstrate the effectiveness of the approach, achieving high sensitivity and accuracy in both lesion-wise Dice and HD95 metrics than the baseline. This scalable method minimizes the dependency on fully annotated data, paving the way for practical real-world applications in clinical settings.
\end{abstract}

\section{Introduction}

Pre-trained GAN model suggested by Ghassemi and others \cite{ghassemi2020deep,braingan} has reach a 98.57\% senstivity to detect a tumor in the BRATS\cite{brats2014} dataset. This makes it possible for us to leverage the high sensitivity of pre-trained GAN-based models to accurately identify and localize tumor regions. By building upon such robust anomaly detection capabilities, we developed advanced segmentation frameworks that refine these initial detections into precise pixel-level tumor boundaries. By "filling in" tumor regions with plausible normal brain structures, the generated masks implicitly capture precise tumor boundaries.

\subsection{Contributions}

Our work combines an anomaly detection network with a mask generation module. The anomaly detection network identifies tumor-sensitive regions, while the mask generation network iteratively refines these regions using sparsity and smoothness constraints to generate high-quality segmentation masks. This framework address the limitations of anomaly detection models, which typically offer high sensitivity coarse localization but lack pixel-level precision. By introducing a mask generation module guided by anomaly detection feedback, our framework refines abnormality localization into precise tumor segmentation. By leveraging abundant unannotated normal brain images and integrating a pre-trained anomaly detection network, our framework minimizes reliance on fully annotated datasets, making it highly scalable and practical for real-world clinical applications.

\section{Background}

The brain is an essential organ in the human body with around one hundred billion neurons \cite{stiles2010brain_development}. Brain tumors refer to abnormal growths that develop within the brain \cite{CROCETTI20121532}. These growths may originate from abnormal cell proliferation in any region of the brain or spine. While the underlying causes of most brain and spinal cancers remain largely unknown, brain tumors are known to manifest a wide range of symptoms due to the diverse functions and regions they can affect. In the United States, brain tumors is the 10th leading cause of cancer-related mortality in 2023, affecting nearly 1 million Americans and contributing to nearly 19,000 deaths annually \cite{siegel2023cancer}.

Accurate segmentation of brain tumors in medical imaging is a critical task that significantly impacts diagnosis, treatment planning, and prognosis. However, the reliance on supervised learning frameworks for tumor segmentation presents a major challenge, as these methods demand large datasets with detailed, pixel-level annotations. The creation of such annotated datasets is time-consuming, labor-intensive, and resource-intensive, limiting the scalability and accessibility of these models in real-world clinical settings.

A significant issue compounding this challenge is the variability and heterogeneity of brain tumors. Tumors differ widely in size, shape, and intensity across imaging modalities, making it difficult for traditional models to generalize effectively \cite{SoTaMRI}. This difficulty is further exacerbated by the imbalance in the availability of labeled data; abnormal brain MRI scans are comparatively scarce, while normal brain MRIs are abundant and easier to collect.

\section{Related Work}

\subsection{Brain Tumor Segmentation}
Traditional segmentation methods, such as atlas-based and region-growing algorithms\cite{hamamci2012tumorcut}, relied on handcrafted features like intensity and texture but struggled with variability and noise. Early CNN-based models processed individual MRI slices with 2D CNNs, offering efficient segmentation for small datasets. The advent of 3D CNNs addressed this limitation, enabling better spatial context and improved segmentation\cite{abidin2024brainsegmentation}.

\subsection{U-Net and Its Advancements}
The U-Net architecture\cite{ronneberger2015unet} remains a cornerstone in medical image segmentation, combining low-level spatial details with high-level semantic information through its encoder-decoder structure with skip connections. Advances such as MultiResUNet\cite{multiresunet}, which integrates multi-resolution blocks and residual connections, and UNet++\cite{unetpp}, with its nested skip pathways, further enhance segmentation accuracy by addressing class imbalance and integrating local and global features.

\subsection{Modalities}
Multi-modal MRI sequences, plays a critical role in brain tumor segmentation by providing complementary tissue-specific information\cite{brats2014}. U-Net variants have effectively utilized these modalities to enhance segmentation accuracy. MM-UNet\cite{mmunet2022} employs separate encoders for each modality, fusing extracted features via attention mechanisms, while Cross-Modality Deep Feature Learning\cite{crossmodality2021} integrates inter-modality relationships for improved tumor delineation. 

\subsection{GANs}
GANs have advanced brain MRI image processing by generating high-quality synthetic images to address data scarcity, improving the robustness of deep learning models. Techniques like Vanilla GANs and DCGANs\cite{ganbrainmri2024} produce realistic MRI scans for dataset augmentation, enhancing tumor detection and anomaly detection by modeling normal tissue distributions. Frameworks like BrainGAN\cite{braingan} combine GAN-generated images with deep learning models for improved segmentation performance. Our proposed model leverages pre-trained GANs for realistic augmentations, boosting sensitivity to tumor regions and refining pixel-level predictions in complex scenarios.

\section{Approach}

\subsection{Model Architectures}
\subsubsection{Pretrained GANs} The GAN pretraining module to learn the is designed to learn the distribution of normal brain from the abundance of unlabeled normal MRI data. The discriminator of this framework is a convolutional network that takes both the generated image and the ground truth normal MRI as input. It evaluates whether the generated image is indistinguishable from the ground truth by classifying it as “real” or “fake”. This discriminator operates on local patches of the image (PatchGAN) to ensure detailed realism in both global structure and local texture. An U-Net based generator is trained adversarially to generate high-quality normal brain images that matches the unmasked surrounding area. After training, the generator is used as a pre-trained module to provide a prior distribution for subsequent segmentation tasks.
\begin{figure}[H]
    \centering
    \begin{subfigure}[b]{0.44\linewidth}
        \centering
        \includegraphics[width=\linewidth]{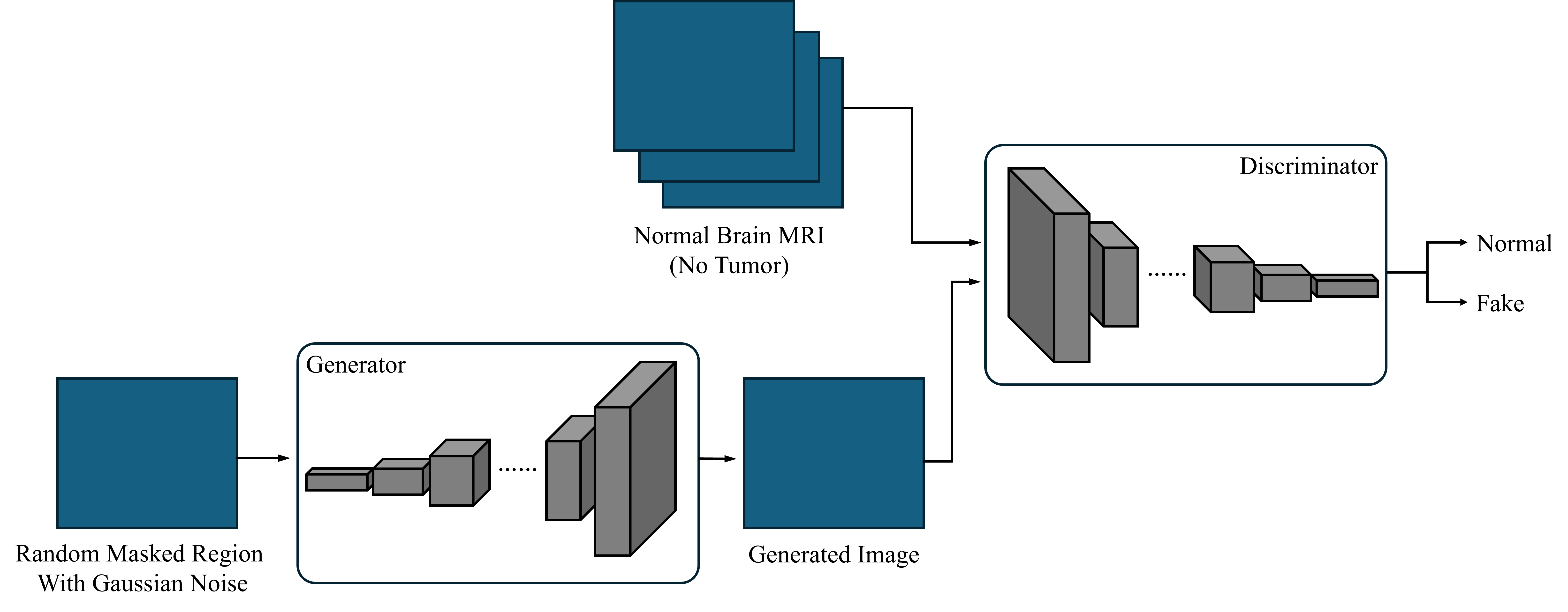}
        \caption{Training Strategy for Brain MRI Reconstruction GANs}
        \label{fig:pretraining1}
    \end{subfigure}
    \hfill 
    \begin{subfigure}[b]{0.54\linewidth}
        \centering
        \includegraphics[width=\linewidth]{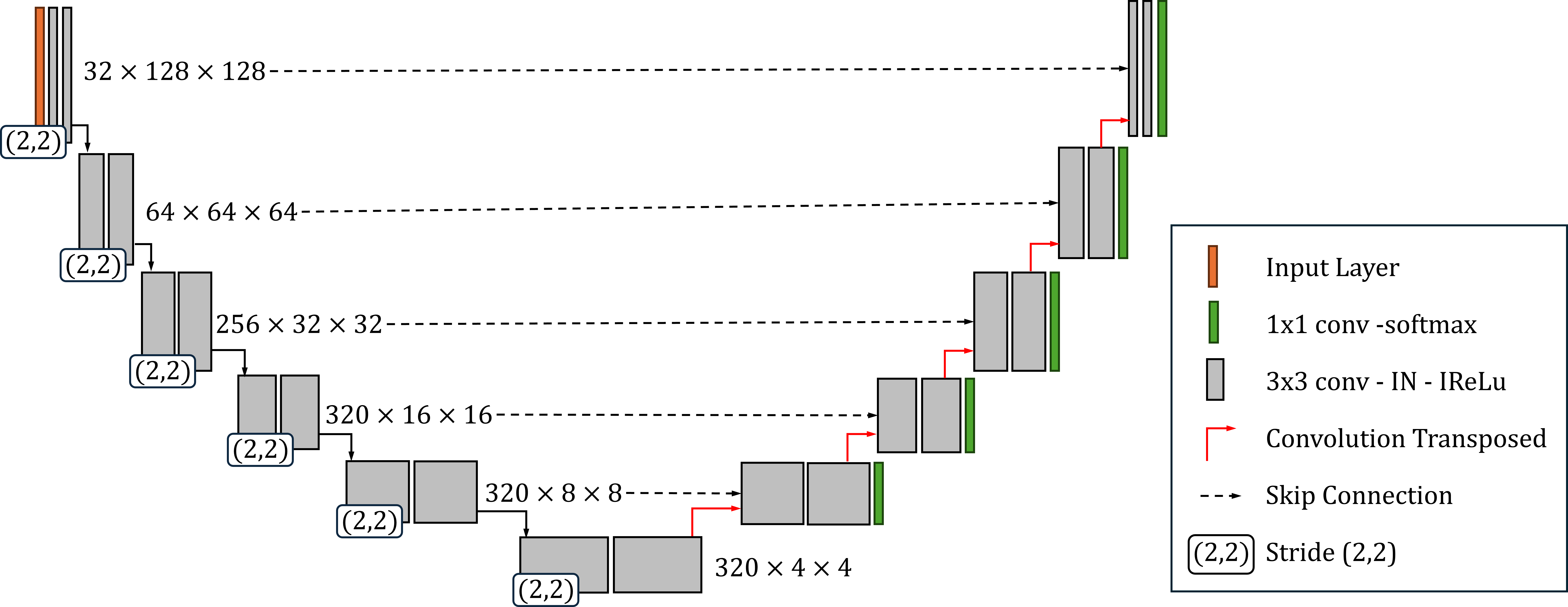}
        \caption{2D Unet Architecture for Multi-Scale Feature Extraction}
        \label{fig:nnunet1}
    \end{subfigure}
    \caption{Training Strategy of GAN model (left) and the Unet \cite{nnunet2021} architecture (right).}
    \label{fig:combined1}
\end{figure}

\subsubsection{Unet} The segmentation module employs an Unet which contains an input layer, an encoder-decoder architecture with skip connections, and a bottleneck to process and segment brain MRI data. The input layer processes the enhanced multi-modality MRI images with flexible dimensions. The encoder consists of multiple convolutional blocks, each with two \(\mathbf{3 \times 3}\) convolutions followed by Instance Normalization (IN) and ReLU activation, alongside down-sampling operations by MaxPooling with \(\mathbf{2 \times 2}\) stride to progressively reduce resolution while extracting high-level semantic features. During the down-sampling process the latent feature channels increasing progressively following \(\mathbf{64 \to 128 \to 256 \to 320}\). The decoder upsamples these features using transposed convolutions, combining them with high-resolution features from the encoder through skip connections to preserve local details. The output layer applies a \(\mathbf{1 \times 1}\) convolution and Softmax activation to produce pixel-wise segmentation probabilities with the same spatial dimensions as the input image. The above figure shows how the segmentation works taking an input of $\mathbf{128 \times 128}$.

\subsubsection{Full Architecture} 
\begin{figure}[H]
    \centering
    \includegraphics[width=0.9\linewidth]{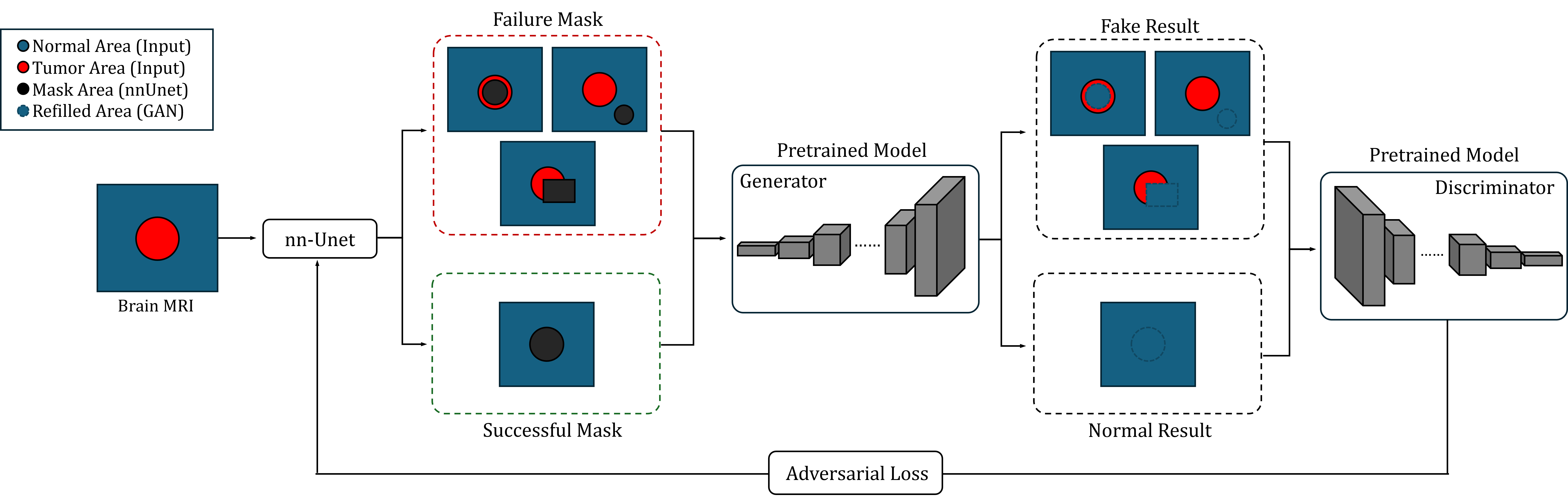}
    \caption{MRI Tumor Segmentation Training Framework with GAN and Unet Integration}
    \label{fig:Full_Arch}
\end{figure}

The model learns the brain tumor segmentation task through a multi-module collaborative optimization framework, as shown in Figure~\ref{fig:Full_Arch}. First, an Unet receives 4 channels modality brain MRI images (T1, T1ce, T2, Flair) as input and generates a mask that indicates regions potentially containing tumors. This mask is then used to occlude the tumor region, and the occluded image is fed into a pre-trained generator. The generator, trained on MRI data of normal brain tissue, reconstructs the input image into a complete normal brain tissue representation by filling in the masked regions as closely as possible to realistic normal tissue. Concurrently, a pre-trained discriminator evaluates the reconstructed image from the generator, determining the probability that each pixels contains abnormal brain tissue (i.e., tumor regions). A $\mathbf{3\times 3}$ Laplacian edge detection kernel and a max-pooling with stride size of $\mathbf{5\times 5}$ are applied to the Unet mask to generate a downsampled edge map with same resolution. The downsampled edge map is dot-multiplied with the output of the pre-trained discriminator, and the output was used to compute adversal loss. 

\section{Experimental Results}

Our experimental setup was carefully designed to ensure optimal performance and accuracy. The hardware infrastructure is based on Amazon EC2 G6 4-GPU instances included 3rd generation AMD EPYC processors, and four NVIDIA L4 Tensor Core GPU each coupled with 22 GB of memory to accelerate computations \cite{aws_ec2_instance_types}. On the software side, the experiments were conducted within a Unbuntu 24.02 environment, and implementation was carried out using Python 3.10 and PyTorch 2.5 framework.

\subsection{Data Pre-processing} 
\subsubsection{Generator and Discriminator} 
The NFBS Repository provides $\mathbf{125}$ T1-weighted MRI scans of healthy brains, each meticulously skull-stripped to align with the BraTS dataset's preprocessing \cite{NFBSReference}. It serves as a source of unlabeled normal brain data for pretraining the GAN module. 

\subsubsection{Segmentation Module}
The BraTS 2020 dataset comprises pre-operative multi-modal MRI scans, including T1, T1Gd, T2, and T2-FLAIR sequences, from patients with glioblastoma (GBM/HGG) and lower-grade glioma (LGG) \cite{MultiModelBrainTumor}. Tumor sub-regions—enhancing tumor (ET), peritumoral edema (ED), and necrotic/non-enhancing tumor core (NCR/NET)—are manually segmented and labeled as 4, 2, and 1, respectively, with healthy tissue labeled as 0 \cite{Bakas2017,bakas2019}. The dataset is preprocessed for co-registration, interpolation to $1 mm^3$ resolution, and skull-stripping. The training set includes $\mathbf{369}$ samples with $\mathbf{1476}$ MRI file, each with $\mathbf{155}$ slices of $\mathbf{240\times240}$ resolution. For benchmarking, $\mathbf{36}$ samples are selected to serve as the test set. These $\mathbf{36}$ samples are used for the final evaluation of the model's performance on unseen data.

\subsection{Evaluation Metrics}
Model performance is assessed using the lesionwise Dice \cite{3dunetlearningdense, ZHANG2021102854} and Hausdorff Distance 95\% (HD95) metrics \cite{Billet_Fedorov_Chrisochoides2008} to compare each region of interest in the ground truth with the corresponding prediction. Lesionwise metrics ensures that all lesions receive equal weight regardless of size, thereby avoiding a bias toward larger lesions \cite{ZHANG2021102854}. The lesionwise Dice computes the overlap for each lesion independently, normalizing the number of true positives to the average size of the two segmented lesion areas. The final score is the mean Dice coefficient across all lesions. The Dice score is mathematically equivalent to the F1 score \cite{MultiModelBrainTumor} and can be monotonously transformed into the Jaccard index. We introduced the lesionwise False negative (FN or sensitivity) and false positive (FP or specificity) predictions rate detailed in Appendix~\ref{app:evaluation}. 

We employed the most commonly used metrics the Hausdorff distance to capture the geometric alignment of the segmentation boundaries. Hausdorff distance (Haus), that evaluates the surface distance between boundaries. A smaller Haus value indicates a smaller maximum mismatch degree between ground-truth and prediction and better segmentation results. We adopt the robust 95-th percentile Hausdorff Distance (HD95) as a standard measure. This metric represents a more reliable variant of the Hausdorff measure by considering the $95\%$ quantile of surface distances and reduces the sensitivity to outliers.

\subsection{Data Augmentation}

The data enhancement pipeline proposed by Ziaur and his fellows \cite{SymmetryBrainTumorSegmentation} has demonstrated significant efficiency in improving the visual quality of MRI images. This method, detailed in Appendix~\ref{app:augmentation}, refines image contrast and brightness through a combination of logarithmic and exponential transformations. Additionally, it employs cumulative distribution function (CDF)-based normalization and linear scaling to enhance the clarity of the images. As shown in Figure 3, the enhanced images exhibit significantly clearer tumor boundaries, making the delineation between the tumor and surrounding tissues more distinct. 

\begin{figure}
    \centering
    \includegraphics[width=0.9\textwidth]{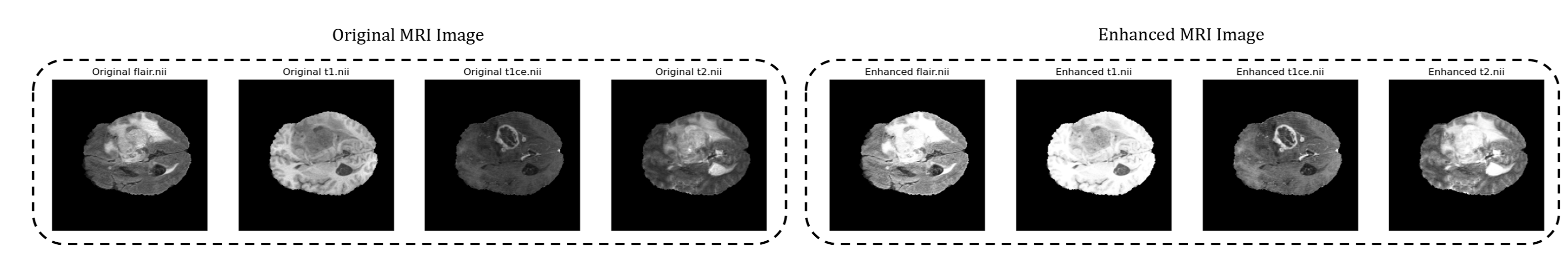}
    \caption{Comparative visualization of original and enhanced MRI images for a single slice across four modalities: T1, T1ce, T2, and FLAIR.}
    \label{fig:enhanced_images1}
\end{figure}

We improved computational efficiency and model performance by cropping each slice of the MRI Image to the smallest rectangular region encompassing the brain with bounding box, as shown in the blue rectangle first figure. As illustrated in the Appendix~\ref{app:figures}, each subject contains slices with blank segmentation masks that contains no useful information for segmentation, and these slices were excluded from the training dataset before training. This approach helps us half the training time per batch from 25.75 seconds to 15.39 seconds on a G6 x4.large server.

\subsection{Hyperparameter Tuning}

\subsubsection{Unet}
The Unet architecture was optimized using the Adam optimizer with an exponentially-decaying learning rate. The initial learning rate was set to $\alpha_0 = 6 \times 10^{-5}$, and subsequent learning rates decayed based on the current epoch relative to the total number of epochs, following a power of $0.75$.

The momentum parameters for Adam were set as $\beta_1 = 0.9$ and $\beta_2 = 0.999$, and weight decay (L2 regularization) was applied with a coefficient of $0.0001$ to prevent overfitting. Training was conducted with a batch size of $80$ over $30$ epochs, balancing memory constraints and gradient stability. A hybrid loss function combining cross-entropy and Dice coefficient losses was used to address class imbalance and ensure accurate segmentation of tumor boundaries. Validation performance was monitored at the end of each epoch using the Dice coefficient, ensuring stable training and improved generalization.

\subsubsection{GAN Model}
The GAN model was trained using the Adam optimizer similar to UNet's for full architecture consistency. Training was conducted over $30$ epochs with a batch size of $80$. To improve training stability, early stopping was employed if validation performance (based on reconstruction loss) stagnated for $5$ consecutive epochs. To avoid discriminator overfitting on the data, the classifier is updated once for every five updates of the generator. 

The generator architecture used ReLU activations in the intermediate layers and Tanh activation in the output layer, ensuring smooth and normalized outputs for reconstructed brain images. The discriminator employed Sigmoid activations in its final layer to output probabilities for pixel-wise classifications (real or fake). Both networks were initialized using Xavier initialization for stable convergence.

\subsection{Results} 
\subsubsection{Sensitivity Analysis on Discriminator}

To analyze the behavior of the discriminator, we adjusted its sensitivity threshold to investigate its impact on tumor detection performance. Table~\ref{tab:sensitivity_results} summarizes the results for different sensitivity thresholds ranging from $0.1$ to $0.4$.

\begin{table}[h]
\centering
\caption{Discriminator Performance Under Different Sensitivity Thresholds}
\begin{tabular}{cccccc}
\toprule
\textbf{Threshold} & \textbf{Accuracy} & \textbf{Sensitivity} & \textbf{True Positives (TP)} & \textbf{False Negatives (FN)} & \textbf{False Positives (FP)} \\
\midrule
0.1 & 75.07\% & 51.24\% & 1219 & 1160 & 231 \\
0.2 & 79.53\% & 88.52\% & 2106 & 273 & 869 \\
0.3 & 76.40\% & 95.00\% & 2260 & 119 & 1198 \\
0.4 & 72.92\% & 98.11\% & 2334 & 45 & 1466 \\
\bottomrule
\end{tabular}
\label{tab:sensitivity_results}
\end{table}

As the threshold increases (e.g., $0.4$), the sensitivity improves dramatically to $98.11\%$, indicating excellent detection of tumor-affected regions. However, as the number of false positives (FP) increases significantly. This behavior is inherently influenced by the nature of Generative Adversarial Networks (GANs), where the generator learns to identify even subtle discrepancies between generative and normal data. The edge-guided adversarial feedback mechanism introduced earlier helps mitigate discriminator overfitting by focusing on boundary discrepancies instead of global variations between generated and true images. 

\subsubsection{Final Results on Tumor Segmentation}

We evaluated the performance of the proposed GANnet-Seg model against baseline segmentation models, including Optimized U-Net, nnU-Net, Swin-UNETR, and UMamba. The results for the baseline models (Optimized U-Net, nnU-Net, Swin-UNETR, and UMamba) were obtained from the paper \textit{``An Ensemble Approach for Brain Tumor Segmentation and Synthesis''}\cite{2024ensembleapproachbraintumor}, which provided a comprehensive comparison of these methods on similar evaluation metrics.

\begin{table}[h]  
    \centering  
    \caption{Final Test Results: Comparison of GANnet-Seg with Baseline Models}  
    \begin{tabular}{lccccccccc}  
        \toprule  
        \textbf{Model} & \multicolumn{3}{c}{\textbf{Dice [\%]}} & \multicolumn{3}{c}{\textbf{HD95 [mm]}} & \multicolumn{3}{c}{\textbf{LW Dice [\%]}} \\
        \cmidrule(lr){2-4} \cmidrule(lr){5-7} \cmidrule(lr){8-10}  
        & \textbf{WT} & \textbf{ET} & \textbf{TC} & \textbf{WT} & \textbf{ET} & \textbf{TC} & \textbf{WT} & \textbf{ET} & \textbf{TC} \\
        \midrule  
        Optimized U-Net  & 81.44 & 72.25 & 75.81 & 22.30 & 48.47 & 29.94 & 68.50 & 67.85 & 71.27 \\
        Swin-UNETR       & 78.59 & 72.51 & 74.75 & 22.85 & 48.83 & 28.90 & 60.96 & 65.75 & 68.03 \\
        UMamba           & \textbf{85.69} & 77.41 & 82.07 & \textbf{14.51} & \textbf{37.48} & 19.68 & 78.03 & 73.95 & 78.05 \\
        \textbf{GANnet-Seg} & 81.28 & \textbf{79.81} & \textbf{88.84} & 27.06 & 50.13 & \textbf{13.95} & \textbf{81.92} & \textbf{74.08} & \textbf{86.28} \\
        \bottomrule  
    \end{tabular}  
    \label{tab:final_results}  
\end{table}

\section{Discussion}
While our proposed framework demonstrates strong performance in brain tumor segmentation, there are notable limitations. The combination of the U-Net segmentation model and the discriminator of the GAN model we developed can accurately identify tumor regions but lacks the ability to delineate the specific tumor sub-regions—\textit{Necrotic and Non-Enhancing Tumor Core (Label 1)}, \textit{Peritumoral Edema (Label 2)}, and \textit{GD-Enhancing Tumor (Label 4)}. This limitation arises because the discriminator, pre-trained on normal brain images, primarily improves sensitivity to tumor detection but does not contribute to the detailed segmentation of these sub-regions, leaving this task solely to the U-Net. Additionally, due to limited computational resources and time, our current approach processes MRI data as 2D slices, which ignores spatial correlations within the 3D volumetric images. Future work will focus on addressing these limitations by extending the U-Net model for multi-class segmentation to differentiate tumor sub-regions while leveraging the GAN’s sensitivity for enhanced refinement. Furthermore, transforming the current framework into a fully 3D segmentation model would allow spatial correlations to be captured effectively, leading to more precise and clinically useful results.

\clearpage
\printbibliography

\clearpage
\appendix

\section{Mathematical Formulation of Loss Functions}  
\label{app:loss}  

The proposed framework integrates multiple loss functions to optimize the GAN and UNet modules collaboratively. The loss functions are introduced in a phased manner to stabilize training, and dynamic weight balancing is employed to adaptively adjust loss contributions.

\subsection{GAN Loss Functions}  

\subsubsection{Discriminator Loss ($\mathcal{L}_D^{\text{GAN}}$):} Ensures the discriminator can distinguish between real and generated images:  
    \begin{equation}
        \mathcal{L}_D^{\text{GAN}} = -\mathbb{E}_{x \sim P_{\text{real}}}[\log D(x)] - \mathbb{E}_{z \sim P_z}[\log(1 - D(G(z)))].
    \end{equation}

\subsubsection{Generator Loss ($\mathcal{L}_G^{\text{GAN}}$):} Encourages the generator to produce images that can deceive the discriminator:  
    \begin{equation}
        \mathcal{L}_G^{\text{GAN}} = -\mathbb{E}_{z \sim P_z}[\log(D(G(z)))].
    \end{equation}

\subsection{Segmentation Loss Functions}  

The UNet segmentation loss combines pixel-wise classification accuracy and spatial overlap precision:

\subsubsection{Cross-Entropy Loss ($\mathcal{L}_{\text{CrossEntropy}}$):} Encourages accurate pixel-wise classification:  
    \begin{equation}
        \mathcal{L}_{\text{CrossEntropy}} = -\sum_{i=1}^N \left[ y_i \log(p_i) + (1 - y_i) \log(1 - p_i) \right],
    \end{equation}
    where \( y_i \) is the ground truth label and \( p_i \) is the predicted probability.  

\subsubsection{Dice Loss ($\mathcal{L}_{\text{Dice}}$)} Improves spatial overlap between predicted masks and ground truth:  
    \begin{equation}
        \mathcal{L}_{\text{Dice}} = 1 - \frac{2 \cdot |P \cap T|}{|P| + |T|},
    \end{equation}
    where \( P \) is the predicted mask and \( T \) is the ground truth mask.

\subsection{Size Consistency and Sparsity Losses}  

\subsubsection{Sparsity Loss ($\mathcal{L}_{\text{Sparsity}}$):} Encourages compact and sparse masks to avoid unnecessary over-segmentation:  
    \begin{equation}
        \mathcal{L}_{\text{Sparsity}} = \alpha \cdot \|M\|_1,
    \end{equation}
    where \( M \) is the predicted mask.

\subsubsection{Size Consistency Loss ($\mathcal{L}_{\text{Size}}$):} Aligns the predicted mask size with the labeled ground truth pixel count \( S_{\text{label}} \):  
    \begin{equation}
        \mathcal{L}_{\text{Size}} = \gamma \cdot \left\| \frac{\sum_{i} M_i - S_{\text{label}}}{S_{\text{label}}} \right\|_1,
    \end{equation}
    where \( S_{\text{label}} \) is the total labeled pixel count from the ground truth, and \( \sum_{i} M_i \) is the predicted mask size.

\subsection{Adversarial Feedback Loss}  

The adversarial feedback loss leverages the discriminator to improve mask precision, particularly at boundaries, by incorporating an edge-focused attention mechanism. Specifically, a Laplacian edge detection operator and mask-based attention are applied:

\begin{equation}
    \mathcal{L}_{\text{Adversarial}} = -\mathbb{E}_{x \sim P_{\text{real}}}[\log D(x)] - \mathbb{E}_{M \sim P_{\text{seg}}}\left[ \log(1 - D(G(x \odot M \cdot E(M)))) \right],
\end{equation}
where \( E(M) \) represents the edge map extracted using a Laplacian filter applied to the mask \( M \), and \( x \odot M \cdot E(M) \) combines the masked input with edge-focused attention. \( D \) is the discriminator, and \( G \) is the generator.

\subsection{Dynamic Weight Balancing and Phased Loss Introduction}

To ensure training stability, the loss weights are introduced in a phased manner:
\subsubsection{Phase 1 (Epoch 0-10):} Only the segmentation loss is optimized:
    \begin{equation}
        \mathcal{L}_{\text{Phase1}} = \mathcal{L}_{\text{Segmentation}}.
    \end{equation}
    
\subsubsection{Phase 2 (Epoch 11+):} The complete loss is introduced, including sparsity, adversarial feedback, and size consistency losses. Dynamic weight balancing adjusts the contribution of each loss based on gradient magnitudes:
    \begin{equation}
        \mathcal{L}_{\text{Total}} = \lambda_{\text{Seg}} \cdot \mathcal{L}_{\text{Segmentation}} + \lambda_{\text{Sparsity}} \cdot \mathcal{L}_{\text{Sparsity}} + \lambda_{\text{Adv}} \cdot \mathcal{L}_{\text{Adversarial}} + \lambda_{\text{Size}} \cdot \mathcal{L}_{\text{Size}},
    \end{equation}
    where the dynamic weights \( \lambda_i \) are defined as:  
    \begin{equation}
        \lambda_i = \frac{1}{\|\nabla \mathcal{L}_i\| + \epsilon},
    \end{equation}
    ensuring that losses with smaller gradients contribute proportionally more to the optimization process.

\section{Edge Refinement with Max-Pooling and Laplacian Filtering}
\label{app:refinement}

The edge refinement process consists of two main steps:

\subsubsection{Max-Pooling Expansion:} The mask $M$ is expanded using a $5 \times 5$ max-pooling operation:
    \begin{equation}
        M_{\text{expanded}}(i, j) = \max_{k, l \in \mathcal{N}_{5 \times 5}(i, j)} M(k, l).
    \end{equation}
\subsubsection{Laplacian Filtering:} The Laplacian filter is applied with a $3 \times 3$ kernel to enhance edge features:
    \begin{equation}
        L(M) = \nabla^2 M = M_{i+1, j} + M_{i-1, j} + M_{i, j+1} + M_{i, j-1} - 4M_{i, j}.
    \end{equation}

This combination refines the boundaries and ensures high-frequency edge details are highlighted for improved segmentation quality.

\section{Mathematical Formulation of the Enhancement Algorithm}
\label{app:augmentation}

The low-complexity image enhancement algorithm by \cite{SymmetryBrainTumorSegmentation} applied the following steps:

\begin{equation}
\text{IMG1} = \frac{\max(I_{\text{img}})}{\log(\max(I_{\text{img}}) + 1)} \times \log(I_{\text{img}} + 1)
\end{equation}
\begin{equation}
\text{IMG2} = 1 - \exp(-I_{\text{img}})
\end{equation}
\begin{equation}
\text{IMG3} = \frac{\text{IMG1} + \text{IMG2}}{\lambda + (\text{IMG1} \times \text{IMG2})}
\end{equation}
\begin{equation}
\text{IMG4} = \text{erf}(\lambda \times \arctan(\exp(\text{IMG3})) - 0.5 \times \text{IMG3})
\end{equation}
\begin{equation}
\text{IMG5} = \frac{\text{IMG4} - \min(\text{IMG4})}{\max(\text{IMG4}) - \min(\text{IMG4})}
\end{equation}

Here, \( I_{\text{img}} \) denotes the initial MRI image, and IMG1 to IMG5 represent intermediate stages of enhancement.

\section{Mathematical Formulation of the Evaluation Metrics}
\label{app:evaluation}
The lesionwise Dice similarity coefficient (DSC) is defined as
\begin{equation}
    {\rm Dice}(P_i,T_i) = {\vert P_i \wedge T_i\vert\over (\vert P_i\vert + \vert T_i\vert)/2}
\end{equation}

where $\wedge$ is the logical AND operator, 
$|\cdot|$ is the number of voxels in the set, and 
$P_i$ and $T_i$ represent the sets of voxels belonging to the i-th lesion in the prediction and ground truth. 

The lesionwise False negative (FN or sensitivity) and false positive (FP or specificity) are defined as
\begin{equation}
    {\rm Sens}(P_i,T_i) = {\vert P_i\wedge T_i\vert\over \vert T_i\vert}\quad {\rm and} \quad {\rm Spec}(P,T) = {\vert P_{i,0} \wedge T_{i,0}\vert \over \vert T_{i,0}\vert}
\end{equation}

where $P_{i,0}$ and $T_{i,0}$ represent the sets of voxels outside the lesion where 
$P=0$ and $T=0$, respectively.

The Hausdorff distance (Haus) defined as
\begin{equation}
    {\rm Haus}(P,T) = \max\{\sup_{p \in \partial P_1} \, \inf_{t \in \partial T_1} ||p,t||,\, \sup_{t \in \partial T_1} \, \inf_{p \in \partial P_1} ||t,p||\}
\end{equation}

\section{Figures and Plots}
\label{app:figures}

\begin{figure}[H]
    \centering
    \includegraphics[width=\linewidth]{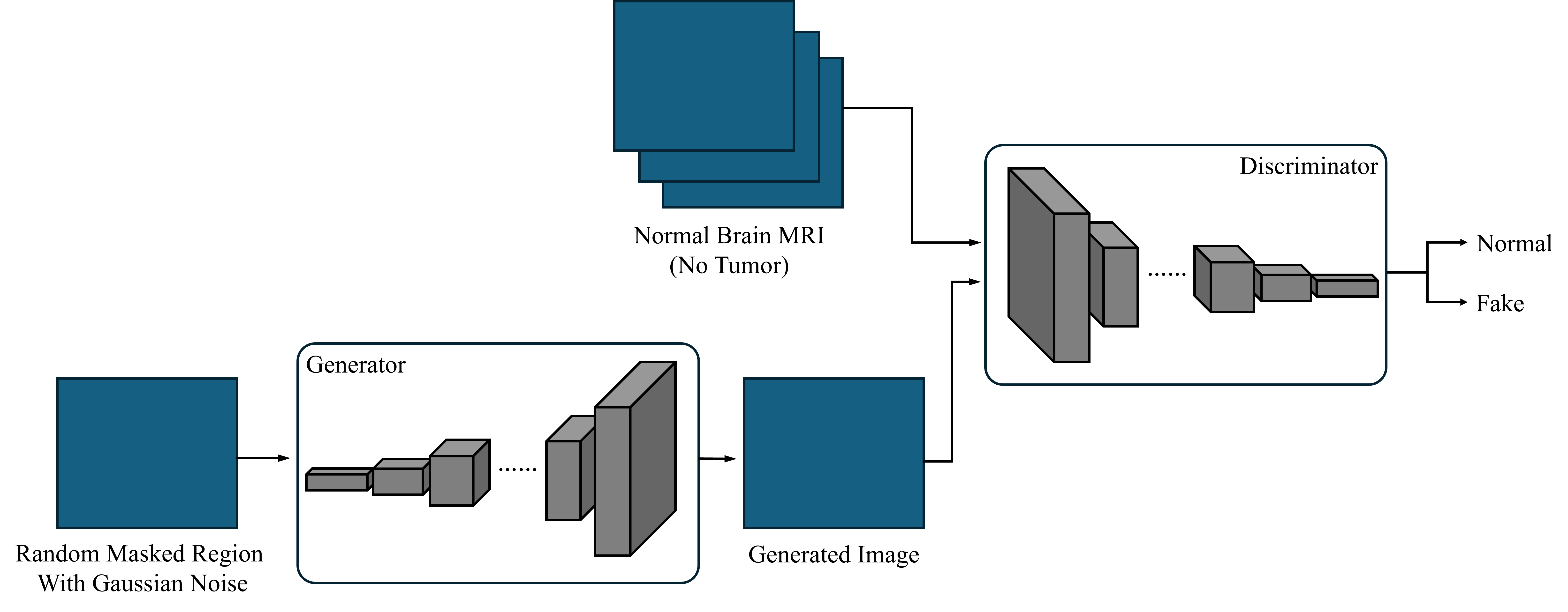}
    \caption{Training Strategy for Brain MRI Reconstruction GANs}
    \label{fig:pretraining2}
\end{figure}

\begin{figure}[H]
    \centering
    \includegraphics[width=\linewidth]{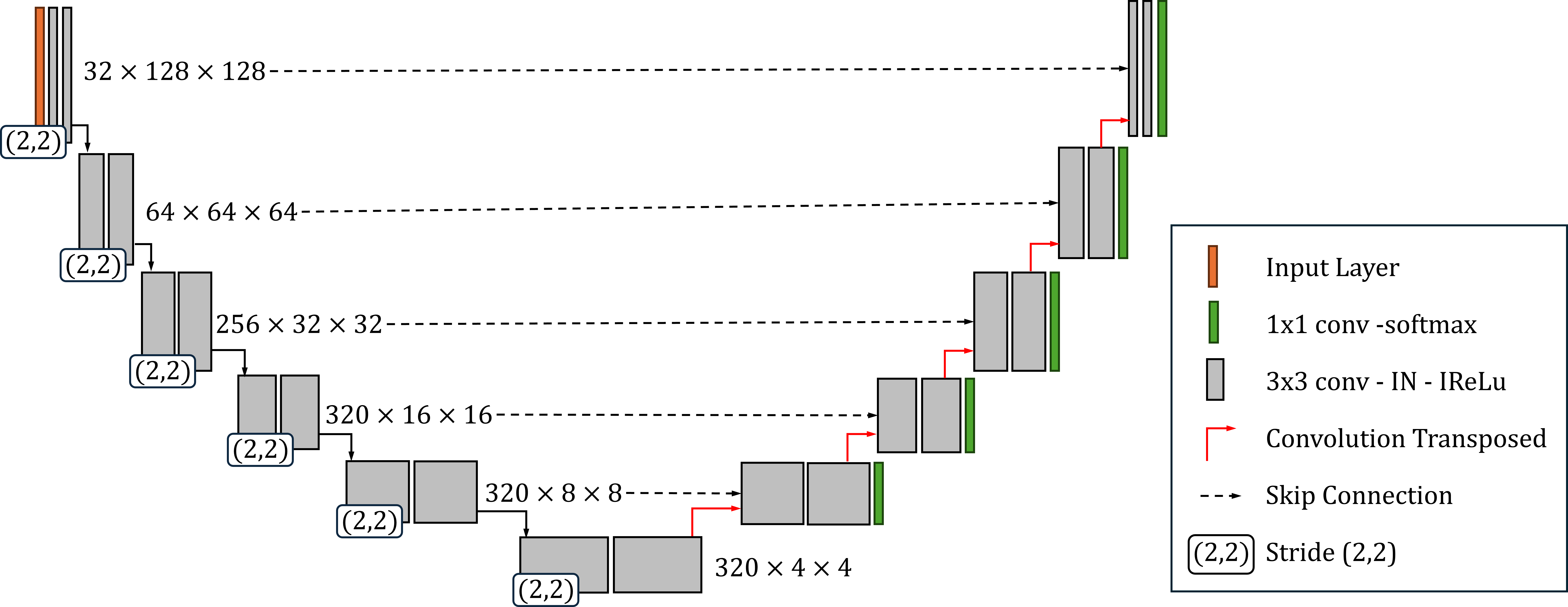}
    \caption{2D Unet Architecture for Multi-Scale Feature Extraction}
    \label{fig:nnunet2}
\end{figure}

\begin{figure}[H]
    \centering
    \includegraphics[width=1\linewidth]{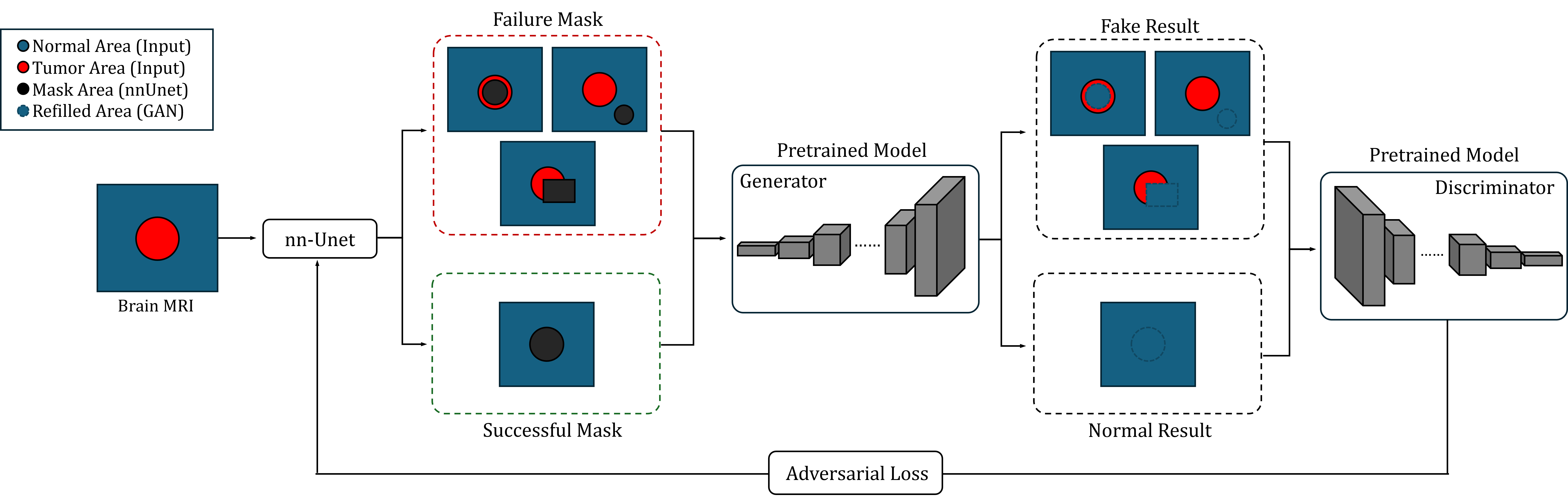}
    \caption{MRI Tumor Segmentation Training Framework with GAN and Unet Integration}
    \label{fig:Full_Arch2}
\end{figure}

\begin{figure}[H]
    \centering
    \includegraphics[width=0.9\textwidth]{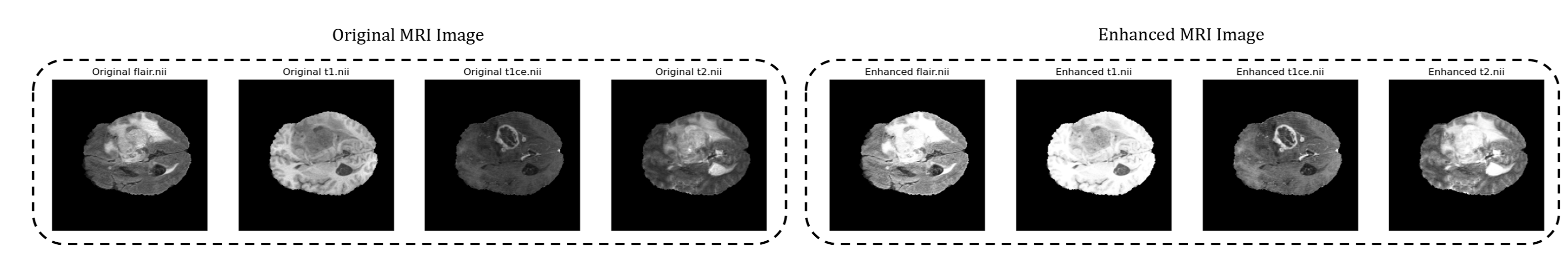}
    \caption{Comparative visualization of original and enhanced MRI images for a single slice across four modalities: T1, T1ce, T2, and FLAIR.}
    \label{fig:enhanced_images2}
\end{figure}

\begin{figure}[H]
    \centering
    \hfill
    \begin{subfigure}[b]{0.33\linewidth}
        \centering
        \includegraphics[width=\linewidth]{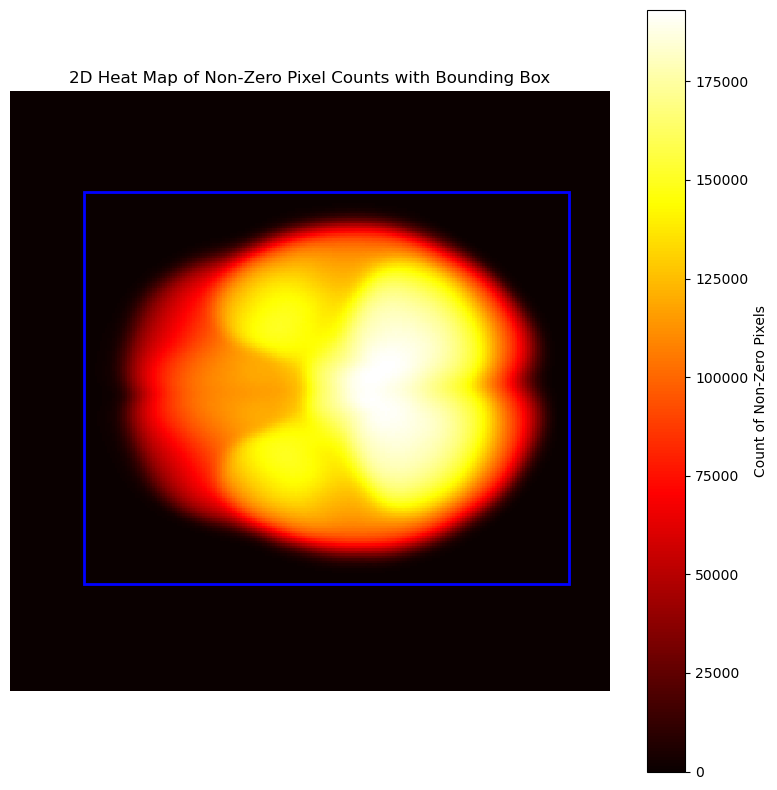}
        \caption{2D Heat Map with Bounding (in Blue)}
        \label{fig:pretraining3}
    \end{subfigure}
    \hfill
    \begin{subfigure}[b]{0.60\linewidth}
        \centering
        \includegraphics[width=\linewidth]{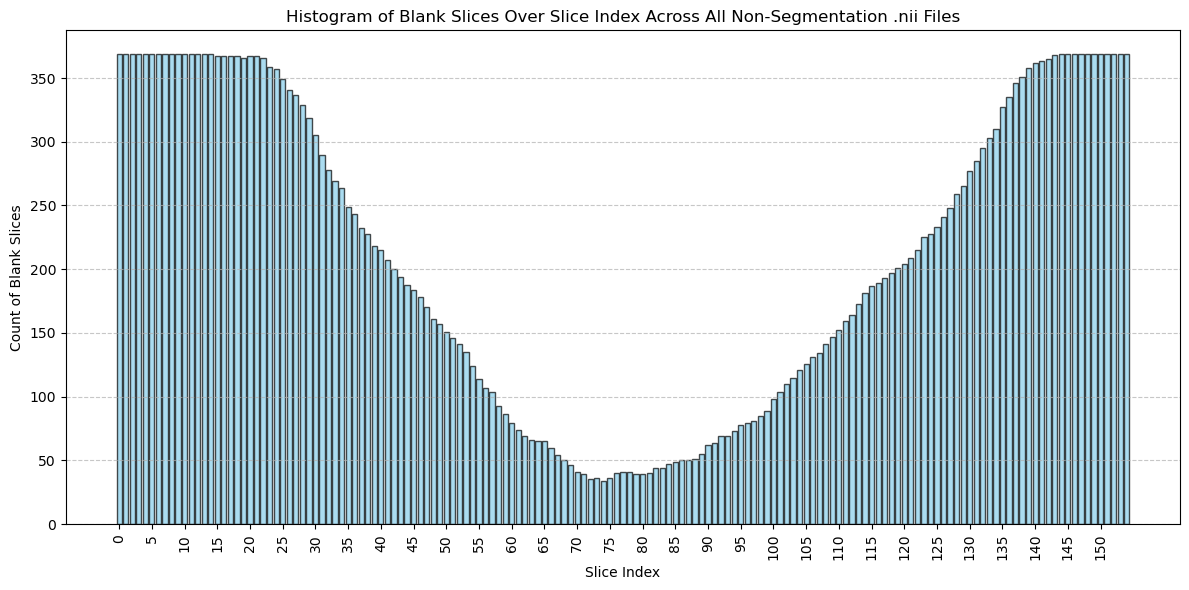}
        \caption{Histogram of Blank Slices Over Slice Index}
        \label{fig:nnunet3}
    \end{subfigure}
    \caption{(Left) 2D heatmap showing the distribution of non-zero pixel counts across slices, with the blue bounding box indicating the smallest rectangular region encompassing the brain. (Right) Histogram of blank slices across all non-segmentation .nii files.}
    \label{fig:combined2}
\end{figure}

\end{refsection}
\end{document}